\documentclass[letterpaper,preprint,prl,superscriptaddress,showpacs]{revtex4}
\usepackage{graphics,graphicx,amsmath}
\usepackage{dcolumn}

\begin{document}
\newcommand{\zl}[2]{$#1\:\text{#2}$}
\newcommand{\zlm}[2]{$#1\:\mu\text{#2}$}
\title{Stopping supersonic oxygen with a series of pulsed electromagnetic coils: A molecular coilgun}
\date{\today}
\author{Edvardas Narevicius}
\affiliation{Center for Nonlinear Dynamics and Department of Physics, The University of Texas at Austin, Austin, Texas 78712-1081, USA}
\author{Adam Libson}
\affiliation{Center for Nonlinear Dynamics and Department of Physics, The University of Texas at Austin, Austin, Texas 78712-1081, USA}
\author{Christian G. Parthey}
\affiliation{Center for Nonlinear Dynamics and Department of Physics, The University of Texas at Austin, Austin, Texas 78712-1081, USA}
\author{Isaac Chavez}
\affiliation{Center for Nonlinear Dynamics and Department of Physics, The University of Texas at Austin, Austin, Texas 78712-1081, USA}
\author{Julia Narevicius}
\affiliation{Center for Nonlinear Dynamics and Department of Physics, The University of Texas at Austin, Austin, Texas 78712-1081, USA}
\author{Uzi Even}
\affiliation{Sackler School of Chemistry, Tel-Aviv University, Tel-Aviv, Israel}
\author{Mark G. Raizen}
\email{raizen@physics.utexas.edu}
\affiliation{Center for Nonlinear Dynamics and Department of Physics, The University of Texas at Austin, Austin, Texas 78712-1081, USA}

\begin{abstract}
We report the stopping of a molecular oxygen beam, using a series of pulsed electromagnetic coils. A series of coils is fired in a timed sequence to bring the molecules to near-rest, where they are detected with a quadrupole mass spectrometer. Applications to cold chemistry are discussed.
\end{abstract}

\pacs{37.10.Mn, 37.20.+j}
\maketitle

The development of new methods to trap and cool molecules paves the way for experimental studies of chemical reactions at ultra-cold temperatures.  In this regime, rate coefficients of exothermic chemical reactions can be determined by the attractive van der Waals interaction and quantum threshold effects.  This is shown, for example, in the benchmark reaction $F + H_2 \rightarrow FH + H$ \cite{Dalgarno}. Another promising application of cold molecules is high-resolution spectroscopy. First experiments on slowed OH radical spectroscopy have been reported \cite{OH_Ye}. \par

Despite the growing interest in the field of ultra-cold chemistry, experimental progress has been hampered by a lack of appropriate methods to trap and cool molecules. Laser cooling, while very successful, is limited to a small number of atoms in the periodic table because few atoms and no molecules have closed cycling transitions. The main methods to produce cold molecules of chemical interest can be divided into two groups. Buffer gas cooling relies on collisions with cold helium in a dilution refrigerator to cool paramagnetic molecules and trap them in a magnetic trap \cite{buffergas}. Supersonic expansion is used by other methods to pre-cool the molecules. The resulting cold molecular beams have been slowed (and trapped in some experiments) by interactions with pulsed electric fields (Stark decelerator) \cite{stark,stark_trap_meijer,stark_trap_ye}, by interactions with pulsed optical fields \cite{barker}, by spinning the nozzle \cite{herschbach}, and by billiard-like collisions \cite{billiard}.  Finally, laser-cooled alkali atoms are used to produce cold molecules via photo-association \cite{photo_Phillips,photo_Heinzen,photo_Stwalley}. None of these methods have, to date, achieved the phase space densities required to observe reaction dynamics at ultra-cold temperatures. \par

We recently demonstrated a general method to stop (and eventually trap) paramagnetic atoms \cite{1stMag,stop_neon}. Our method is based on the interaction of a paramagnetic particle with pulsed magnetic fields. It operates in analogy with the Stark decelerator \cite{stark,stark_trap_meijer,stark_trap_ye} by reducing the kinetic energy of a paramagnetic atom passing through a series of pulsed electromagnetic coils. The amount of kinetic energy removed by each stage is equal to the Zeeman energy shift that the atom experiences at the time the fields are being switched off. The same approach has been independently pursued \cite{merkt1,merkt2}. Our method relies on the presence of a permanent magnetic moment and may be applied to any paramagnetic species that can be entrained or seeded into a beam. As such, the coilgun can be used to slow and eventually trap paramagnetic molecules. Most molecular radicals have a magnetic moment in the ground electronic state, and several molecules such as $NO$, $NO_2$, and $ClO_2$ have a non-zero electron spin in the ground electronic state which means that a substantial fraction of these molecules will be paramagnetic at supersonic expansion temperatures.  Molecular oxygen is a unique case, as the ground state is itself paramagnetic. \par

In this Letter we demonstrate the stopping of ground state $^3\Sigma_{g}^{-}$ oxygen using a 64 stage coilgun. We slow oxygen molecules from \zl{389}{m/s} to \zl{83}{m/s}, removing over $95 \%$ of the initial kinetic energy. The final velocity was only limited by the spread of the beam between the last coil and the detector. We chose oxygen because it has a permanent magnetic moment in the ground electronic state and is important to chemical processes such as combustion reactions, surface interactions, and atmospheric chemistry. Slowing of molecules is complicated by the existence of avoided energy level crossings between different rotational states at high magnetic fields. In light of this complexity, an experimental realization of magnetic stopping of molecules is essential in order to establish the viability of the method. \par

We now discuss the magnetic sub-levels involved in magnetic deceleration of oxygen. Nuclear statistics forbid the existence of $K=0,2,4,\ldots$ rotational levels of  $^{16}O_2$. The addition of electron spin angular momentum, $S$, to the rotational angular momentum, $K$, gives the three possible total angular momentum states $J=0,1,2$ ($J=K+S$, electron orbital angular momentum in the $\Sigma$ state is zero). In the weak-field Zeeman approximation, the three total angular momentum states split into 9 sub-levels with different total angular momentum projections on the quantization axis. At high fields (above \zl{2.5}{T}) the electron spin decouples from the rotational angular momentum and forms three sets of three levels, each set nearly-degenerate. The $J=2$, $M_J=2$ sub-level has the highest magnetic moment both in the low and high field regions, where the magnetic moment is approximately equal to 1.8 Bohr magnetons \cite{oxygen_moment} in the Paschen-Back regime. At about \zl{8}{T} the $J=2$, $M_J=2$ sub-level mixes with the $J=2$, $M_J=2$ sub-level from the $K=3$ manifold, changing its character from low-field to high-field seeking. Since the high-field seeking state is defocused by the fields created in our solenoids, this sets a limit on the maximal value of the magnetic fields that we can use. The measured peak value of the magnetic fields created on the axis of our solenoids is \zl{5.2}{T}. Based on our finite element calculations, this corresponds to a maximal magnetic field adjacent to the coil windings of just below \zl{6}{T} which is well below the limit of \zl{8}{T}. As such, we do not expect any losses due to the avoided crossing. In order to achieve the highest slowing efficiency, we tune the timing of our coils such that molecules in the ground rotational $J=2$, $M_J=2$ sub-level are slowed. \par
\begin{figure}
\includegraphics[width=0.48\textwidth]{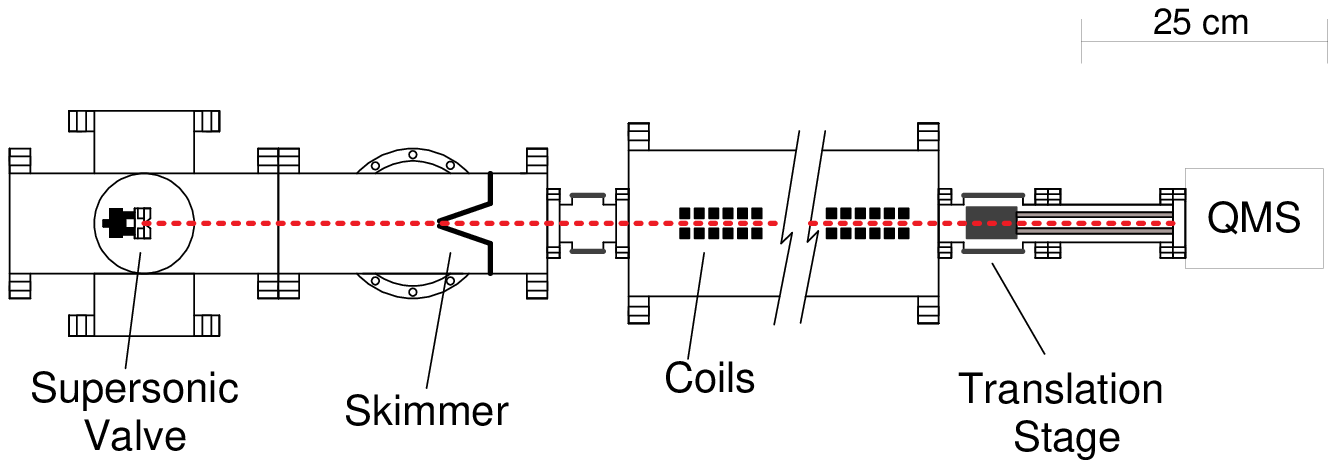}
\caption{\label{fig1} A schematic drawing of our apparatus. The center section has been shorted by a \zl{64}{cm} cutout. Objects other than the coils are to scale; coils have been enlarged for clarity.}
\end{figure}
\begin{figure}
\includegraphics[width=0.5\textwidth]{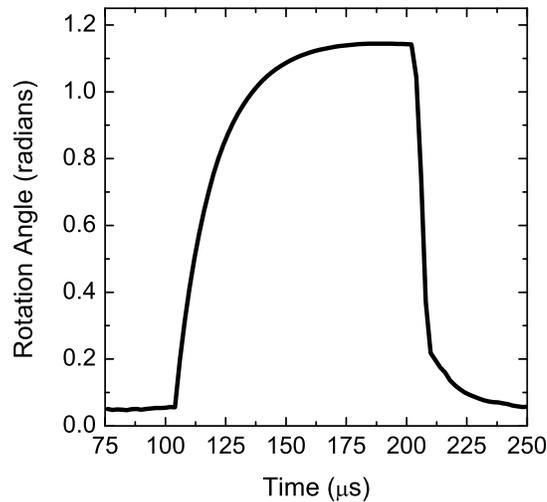}
\caption{\label{fig2} Rotational angle of linearly polarized HeNe laser passing through a 1.4 mm TGG crystal placed in the bore of a pulsed coil as a function of time. The curve is an average of 200 individual measurements.}
\end{figure}
We now describe our experimental setup, which is illustrated in Fig. \ref{fig1}. We generate the supersonic beam using the Even-Lavie pulsed solenoid valve \cite{uzi1,uzi2}. It is capable of producing  \zlm{10}{s} FWHM pulses with intensities of up to \zl{10^{24}\:}{atoms/sr/s}. In order to lower the initial velocity of the beam we use a mixture of oxygen and krypton at a ratio of $1 \colon 5$, and cool the valve to \zl{148 \pm 1}{K}. This produces an oxygen beam with a mean velocity of \zl{389}{m/s} with \zl{28}{m/s} standard deviation. The beam passes through a \zl{5}{mm} diameter, \zl{50}{mm} long, conical skimmer which is mounted \zl{300}{mm} from the exit of the nozzle. From the skimmer base the beam travels  \zl{250}{mm} to the first coil of our slower. \par 

We use the same slower design as was used to stop the metastable neon beam \cite{stop_neon}. It consists of 64, \zl{14}{mm} spaced (center-to-center) electromagnetic coils (similar in design to the solenoid used in our pulsed valve) that are controlled individually by passing a peak current of \zl{750}{A} in a \zlm{100}{s} long pulse through 30 copper windings (\zl{0.5}{mm} wire diameter). Due to the high magnetic saturation materials that we use to confine the magnetic flux we achieve \zl{5.2}{T} peak magnetic fields on the axis of our solenoid. We switch the magnetic field to 20 \% of its initial value in \zlm{6}{s}. The remaining field, due to the eddy currents, decays exponentially with a time constant of \zlm{17}{s}. In order to characterize the switching profile and absolute magnetic field values we insert a magneto-optical crystal, terbium gallium garnet (TGG), into the bore of our coil and shine a linearly polarized HeNe laser beam focused to \zlm{80}{m} along the axis. We monitor the polarization direction of this beam throughout the switching process \cite{faraday}. The polarization of light propagating in a magneto-optical material rotates by an amount proportional to the magnetic field parallel to the light (Faraday effect). The polarization rotation profile of a pulsed coil is shown in Fig. \ref{fig2} for a TGG crystal length of \zl{1.4}{mm}. \par
\begin{figure}
\includegraphics[width=0.5\textwidth]{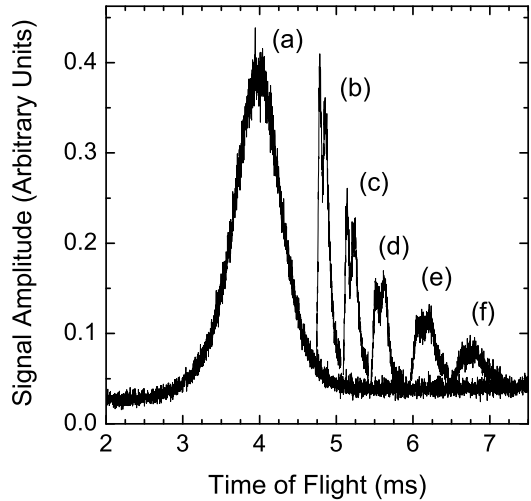}
\caption{\label{fig3} Time-of-flight measurements recorded with the QMS detector for different phases along with a reference beam. The slowed curves show only the slowed portion of the beam for clarity. Beam velocities are (a) \zl{389}{m/s}, (b) \zl{242}{m/s}, (c) \zl{195}{m/s}, (d) \zl{155}{m/s}, (e) \zl{114}{m/s}, and (f)  \zl{83}{m/s}. Each profile is an average of 200 individual measurements.}
\end{figure}
\begin{figure}[t]
\includegraphics[width=0.5\textwidth]{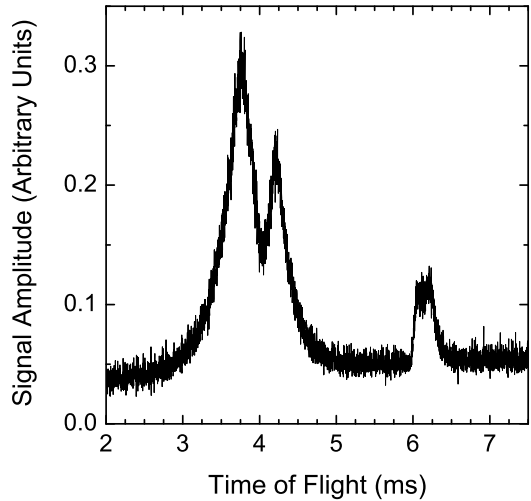}
\caption{\label{fig4} Full range time-of-flight measurement recorded with the QMS detector for \zl{114}{m/s} final velocity. This figure shows the perturbed initial beam along with the slowed peak. The curve is an average of 200 individual measurements.}
\end{figure}
We detect the oxygen beam using a quadrupole mass spectrometer (QMS) equipped with an electron multiplier output \cite{RGA}. The detector is mounted on a \zl{50.8}{mm} translation stage, which allows us to directly measure the velocity of the beam. The distance between the last decelerator coil and the ionizer is \zl{100}{mm}. \par

The parameter we adjust to control the final velocity of the beam is the position of a synchronous molecule inside of a coil when it is switched off.  We follow the convention of the Stark decelerator \cite{meijer-phase} and refer to this position as a phase angle. Phase stable acceleration has its roots in particle accelerator physics \cite{accelerator1, accelerator2}. A phase angle of $0^{\circ}$ corresponds to switching the coil when a synchronous molecule is between two coils, while $90^{\circ}$ refers to switching when the molecule is at the center of the coil. A lower switching phase means that a molecule experiences lower peak magnetic fields and the molecule loses less kinetic energy per deceleration stage.  However, switching at lower phase increases the volume of the stable phase space region and leads to greater fluxes of slowed molecules. The switching sequence is generated by numerically simulating the flight of a molecule through our apparatus using magnetic field gradients calculated by finite element analysis and a temporal profile obtained by Faraday rotation measurement. Since the deceleration process is adiabatic it does not depend strongly on the fine details of the switching profile. Even changing the magnetic moment value by as much as $10 \%$ when numerically calculating the timing sequence and keeping the terminal velocity constant by adjusting the phase, we do not detect changes in the slow molecule flux. \par

A possible loss mechanism for a pulsed magnetic slower is Majorana spin flips at zero magnetic field \cite{merkt1,merkt2}. In our case, during the initial deceleration stages the adjacent coil current pulses overlap. It takes about \zlm{35}{s} for a molecule traveling at \zl{390}{m/s} to cross a single solenoid. Since our pulse length is fixed at \zlm{100}{s} there is a \zlm{65}{s} overlap in adjacent coil pulses. For the slowest molecules traveling at \zl{80}{m/s}, the time between adjacent coils switching on increases to \zlm{170}{s} leaving a \zlm{70}{s} time interval were no current pulse is applied. However, due to the slowly decaying residual field (generated by eddy currents) molecules always experience a finite magnetic field and we do not expect Majorana losses in our deceleration process. \par

The results of our experiments are shown in Fig. \ref{fig3}.  We show slowing of a molecular oxygen beam operating our coilgun with phase angles between $47.8^{\circ}$ and $63.2^{\circ}$.  The initial beam has a velocity of \zl{389\pm 5}{m/s} with a standard deviation of \zl{28}{m/s}.  The slowed peaks have velocities of $242\pm 13$, $195\pm 8$, $155\pm 5$, $114\pm 3$, and \zl{83 \pm 3}{m/s} respectively.  For the \zl{83}{m/s} beam, more than 95 \% of the kinetic energy is removed.  The progression of the slowed peaks for different switching phases demonstrates that this method of slowing is tunable and can be used to create beams of any velocity.  The shape of the slowed peak is determined by the phase used, as well as the profile of the magnetic fields in the coils.  The anharmonic nature of the potential a molecule experiences as it enters a coil leads to different periods for different phase space orbits.  Thus, the time of flight profile of the slowed beam is not expected to be a single peak.  As can be seen in Fig. \ref{fig4}, we do not slow the entire initial beam, but instead slow molecules in a particular velocity window.  The size of this window depends on the phase angle used by the slower. According to our calculations the window is approximately \zl{\pm 2}{m/s} for the highest phase used. Also, molecules which are high-field seekers will be defocused by the magnetic fields in the coils, and are thus expected to be largely removed from the beam.  This contributes to the minimum seen in the perturbed initial beam. \par

We now provide criteria for the general application of the coilgun to stopping paramagnetic molecules. As mentioned above, avoided crossings may limit the strength of the field that can be applied to a low-field seeking molecule. The maximal field strength can be estimated knowing the rotational constant and magnetic moment of the molecule. For example in the case of molecular oxygen, $^{16}O_2$, the difference between the lowest $(K=1)$ and the first excited $(K=3)$ rotational states is equal to $10B$, where $B=43100$ MHz is the rotational constant. Since the maximum magnetic moment of ground state molecular oxygen is about 2 Bohr magnetons (corresponding to 2 unpaired electrons), the energy splitting between the two rotational states in the Paschen-Back regime decreases as $2 \cdot 2\mu_B H$, where $H$ is magnetic field and $\mu_B$ is the Bohr magneton. The two levels cross at the magnetic field value that satisfies the equation, $2 \cdot 2\mu_B H_{max}=10B$. Solving for $H_{max}$ gives a magnetic field value of \zl{7.7}{T}. This field value of the avoided crossing imposes a maximum energy that can be removed from the beam in a single slowing stage. As such, a design must allow for a large number of deceleration stages in order to slow heavy molecules. Our design possesses the required scalability, as we have demonstrated by extending our apparatus from 18 stages \cite{1stMag} to 64 stages \cite{stop_neon}. \par 

An alternative solution to the problem of avoided crossings would be to construct a decelerator for diamagnetic species and high-field seeking states based on an alternating gradient method, as has been shown in the case of the Stark decelerator \cite{AG_Stark}. Here, the magnetic field strength limitation is removed and higher fields would enable more efficient slowing. Another possibility is to trap low-field seeking molecules in a traveling magnetic trap,  similar to the atom ``conveyor'' used to transport cold atoms over macroscopic distances \cite{Hansch}, and decelerate the trapped molecules adiabatically \cite{TheoMag}. One could choose a low acceleration such that trapped molecules would not substantially deviate from the field minimum at the center of the magnetic trap. \par

Future work will extend the coilgun method to trap molecules of chemical interest. Once they are in a magnetic trap, they can be cooled to near the single photon recoil limit using the method of single photon cooling, as demonstrated recently with trapped atoms \cite{SPACth,SPACexp}. The application of this method to cooling of molecules is particularly promising \cite{SPMC_OH}. \par

This work is supported by the  Army Research Office, the R.A.~Welch Foundation, the Sid W. Richardson Foundation  and the US National Science Foundation Physics Frontier Center.\par

\end{document}